\documentclass[aps,prb,twocolumn,showpacs,superscriptaddress,longbibliography]{revtex4-2}
\usepackage{amsfonts}
\usepackage{amsmath}
\usepackage{graphicx}
\usepackage{bm}
\usepackage{amssymb}
\usepackage{dcolumn}
\usepackage{color}
\usepackage{multirow}
\usepackage{booktabs}
\usepackage[colorlinks,
linkcolor=blue,
anchorcolor=blue,
citecolor=blue,
urlcolor=blue,
]{hyperref}

\begin{document}

\title{Two-dimensional Heisenberg models with materials-dependent superexchange interactions}

\author{Jia-Wen Li}
\affiliation{Kavli Institute for Theoretical Sciences, University of Chinese Academy of Sciences, Beijng 100049, China}

\author{Zhen Zhang}
\affiliation{Key Laboratory of Multifunctional Nanomaterials and Smart Systems, Division of Advanced Materials, Suzhou Institute of Nano-Tech and Nano-Bionics, Chinese Academy of Sciences, Suzhou, 215123 China}

\author{Jing-Yang You}
\affiliation{Department of Physics, National University of Singapore, Science Drive, Singapore 117551}
\author{Bo Gu}
\email{gubo@ucas.ac.cn}
\affiliation{Kavli Institute for Theoretical Sciences, University of Chinese Academy of Sciences, Beijng 100049, China}
\affiliation{CAS Center for Excellence in Topological Quantum Computation, University of Chinese Academy of Sciences, Beijng 100190, China}
\affiliation{Physical Science Laboratory, Huairou National Comprehensive Science Center, Beijing 101400, China}

\author{Gang Su}
\email{gsu@ucas.ac.cn}
\affiliation{Kavli Institute for Theoretical Sciences, University of Chinese Academy of Sciences, Beijng 100049, China}
\affiliation{CAS Center for Excellence in Topological Quantum Computation, University of Chinese Academy of Sciences, Beijng 100190, China}
\affiliation{Physical Science Laboratory, Huairou National Comprehensive Science Center, Beijing 101400, China}
\affiliation{School of Physical Sciences, University of Chinese Academy of Sciences, Beijng 100049, China}

\begin{abstract}
The two-dimensional (2D) van der Waals ferromagnetic semiconductors, such as CrI$_3$ and Cr$_2$Ge$_2$Te$_6$, and the 2D ferromagnetic metals, such as Fe$_3$GeTe$_2$ and MnSe$_2$, have been obtained in recent experiments and attracted a lot of attentions. The superexchange interaction has been suggested to dominate the magnetic interactions in these 2D magnetic systems. In the usual theoretical studies, the expression of the 2D Heisenberg models were fixed by hand due to experiences. Here, we propose a method to determine the expression of  the 2D Heisenberg models by counting the possible superexchange paths with the density functional theory (DFT) and Wannier function calculations. With this method, we obtain a 2D Heisenberg model with six different nearest-neighbor exchange coupling constants for the 2D ferromagnetic metal Cr$_3$Te$_6$, which is very different for the crystal structure of Cr atoms in Cr$_3$Te$_6$. The calculated Curie temperature Tc = 328 K is close to the Tc = 344 K of 2D Cr$_3$Te$_6$ reported in recent experiment. In addition, we predict two stable 2D ferromagnetic semiconductors Cr$_3$O$_6$ and Mn$_3$O$_6$ sharing the same crystal structure of Cr$_3$Te$_6$. The similar Heisenberg models are obtained for 2D Cr$_3$O$_6$ and Mn$_3$O$_6$, where the calculated Tc is 218 K and 208 K, respectively. Our method offers a general approach to determine the expression of Heisenberg models for these 2D magnetic semiconductors and metals, and builds up a solid basis for further studies.

\end{abstract}
\maketitle

\section{Introduction}
Recently, the successful synthesis of two-dimensional (2D) van der Waals ferromagnetic semiconductors in experiments, such as CrI$_3$ \cite{Huang2017} and Cr$_2$Ge$_2$Te$_6$  \cite{Gong2017} has attracted extensive attentions to 2D ferromagnetic materials. According to Mermin-Wagner theorem \cite{Mermin1966a}, the magnetic anisotropy is essential to produce the long-range magnetic order in 2D systems. For the 2D magnetic semiconductors obtained in experiments, the Curie temperature Tc is still much lower than room temperature. For example, Tc = 45 K in CrI$_3$ \cite{Huang2017}, 30 K in  Cr$_2$Ge$_2$Te$_6$ \cite{Gong2017}, 34 K in CrBr$_3$ \cite{Zhang2019}, 17 K  in CrCl$_3$ \cite{Cai2019}, 75 K in Cr$_2$S$_3$ \cite{Cui2019,Chu2019}, etc. For applications, the ferromagnetic semiconductors with Tc higher than room temperature are highly required \cite{Zhao2022,Zhao2021a,Huang2021,Sun2020a}. On the other hand, the 2D van der Waals ferromagnetic metals with high Tc have been obtained in recent experiments. For example, Tc = 140 K in CrTe \cite{Wang2020}, 300 K in CrTe$_2$ \cite{Meng2021,Zhang2021a}, 344 K in Cr$_3$Te$_6$ \cite{Chua2021}, 160 K in Cr$_3$Te$_4$ \cite{Li_2022}, 280 K in CrSe \cite{Zhang2019b}, 300 K in Fe$_3$GeTe$_2$ \cite{Fei2018,Deng2018}, 270 K in Fe$_4$GeTe$_2$ \cite{Seo2020}, 229 K in Fe$_5$GeTe$_2$ \cite{Chen2022,May2019}, 300 K in MnSe$_2$ \cite{Ohara2018}, etc. 

In these 2D van der Waals ferromagnetic materials, the superexchange interaction has been suggested to dominate the magnetic interactions.  The superexchange interaction describes the indirect magnetic interaction between two magnetic cations mediated by the neighboring non-magnetic anions  \cite{Anderson1950,Goodenough1955,Kanamori1957}. The superexchange interaction has been discussed in the 2D magnetic semiconductors.
Based on the superexchange interaction, the strain-enhanced Tc in 2D ferromagnetic semiconductor Cr$_2$Ge$_2$Se$_6$ can be understood by the decreased energy difference between the d electrons of cation Cr atoms and the p electrons of anion Se atoms \cite{Dong2019}. The similar superexchange picture was obtained in several 2D ferromagnetic semiconductors, including the great enhancement of Tc in bilayer heterostructures Cr$_2$Ge$_2$Te$_6$/PtSe$_2$ \cite{Dong2020}, the high Tc in technetium-based semiconductors TcSiTe$_3$, TcGeSe$_3$ and TcGeTe$_3$ \cite{You2020}, and the electric field enhanced Tc in the monolayer MnBi$_2$Te$_4$ \cite{You2021}. The superexchange interaction has also been discussed in the semiconductor heterostructure CrI$_3$/MoTe$_2$ \cite{Chen2019a}, and 2D semiconductor Cr$_2$Ge$_2$Te$_6$ with molecular adsorption \cite{He2019}. 

In addition, the superexchange interaction has also been obtained in the 2D van der Waals ferromagnetic metals. By adding vacancies, the angles of the superexchange interaction paths of 2D metals VSe$_2$ and MnSe$_2$ will change, thereby tuning the superexchange coupling strength \cite{Li2022a}. It is found that biaxial strain changes the angle of superexchange paths in 2D metal Fe$_3$GeTe$_2$, and affects Tc \cite{Hu2020}. Under tensile strain, the ferromagnetism of the 2D magnetic metal CoB$_6$ is enhanced, due to the competition between superexchange and direct exchange interactions \cite{Tang2019}.

It is important to determine the spin Hamiltonian for the magnetic materials, in order to theoretically study the magnetic properties, such as Tc. In the usual theoretical studies, the expression of the spin Hamiltonian needs to be fixed by hand according to the experiences. By the four-state method and density functional theory (DFT) calculations  \cite{Xiang2013,Xiang2011,Li2021a}, the exchange coupling parameters of the spin Hamiltonian, such as the nearest neighbor, the next nearest neighbor, inter-layer, etc, can be obtained. Then the Tc can be estimated through Monte Carlo simulations \cite{Li2021a}. With different spin Hamiltonians chosen by hand, sometimes different results are  obtained in calculations. Is it possible to determine the spin Hamiltonian by the help of calculations rather than by the experiences ?

In this paper, we propose a method to establish the 2D Heisenberg models for the 2D van der Waals magnetic materials, when the superexchange interactions dominate. Through the DFT and Wannier function calculations, we can calculate the exchange coupling between any two magnetic cations, by counting the possible superexchange paths. By this method, we obtain a 2D Heisenberg model with six different nearest-neighbor exchange coupling constants for the 2D van der Waals ferromagnetic metal Cr$_3$Te$_6$ \cite{Chua2021}, where the calculated Tc = 328 K is close to the Tc = 344 K reported in the experiment. In addition, based on the crystal structure of 2D Cr$_3$Te$_6$, we predict two 2D magnetic semiconductors Cr$_3$O$_6$ and Mn$_3$O$_6$ with Tc of 218 K and 208 K, and energy gap of 0.99 eV and 0.75 eV, respectively.

\section{Computational methods}
Our calculations were based on the DFT as implemented in the Vienna ab initio simulation package (VASP) \cite{Kresse1996a}. The exchange-correlation potential is described with the Perdew-Burke-Ernzerhof (PBE) form of the generalized gradient approximation (GGA) \cite{Perdew1996a}. The electron-ion potential is described by the projector-augmented wave (PAW) method \cite{Bloechl1994a}. We carried out the calculation of GGA + U with U = 3.2 eV, a reasonable U value for the 3d electrons of Cr in Cr$_3$Te$_6$  \cite{Chua2021}. The band structures for 2D Cr$_3$O$_6$ and Mn$_3$O$_6$ were calculated in HSE06 hybrid functional \cite{Heyd2003}. The plane-wave cutoff energy is set to be 500 eV. Spin polarization is taken into account in structure optimization. To prevent interlayer interaction in the supercell of 2D systems, the vacuum layer of 16 Å is included. The 5$\times$9$\times$1, 5$\times$9$\times$1 and 7$\times$11$\times$1 Monkhorst Pack k-point meshed were used for the Brillouin zone (BZ) sampling for 2D Cr$_3$O$_6$, Cr$_3$Te$_6$ and Mn$_3$O$_6$, respectively \cite{Monkhorst1976a}. The structures of 2D Cr$_3$O$_6$ and Mn$_3$O$_6$ were fully relaxed, where the convergence precision of energy and force were $10^{-6}$ and $10^{-3}$ eV/Å, respectively. The phonon spectra were obtained in a 3×3×1 supercell with the PHONOPY package \cite{Togo2015}. The Wannier90 code was used to construct a tight-binding Hamiltonian \cite{Mostofi2014a,Mostofi2008} to calculate the magnetic coupling constant. In the calculation of molecular dynamics, a 3$\times$4$\times$1 supercell (108 atoms) was built, and we took the NVT ensemble (constant-temperature, constant-volume ensemble) and maintained a temperature of 250 K with a step size of 3 fs and a total duration of 6 ps.

\section{M\lowercase{ethod to determine the 2\uppercase{D} \uppercase{H}eisenberg model: An example of 2\uppercase{D} \uppercase{C}r$_3$\uppercase{T}e$_6$}}

\begin{figure*}[!htbh]
	\centering
	\includegraphics[scale=0.6,angle=0]{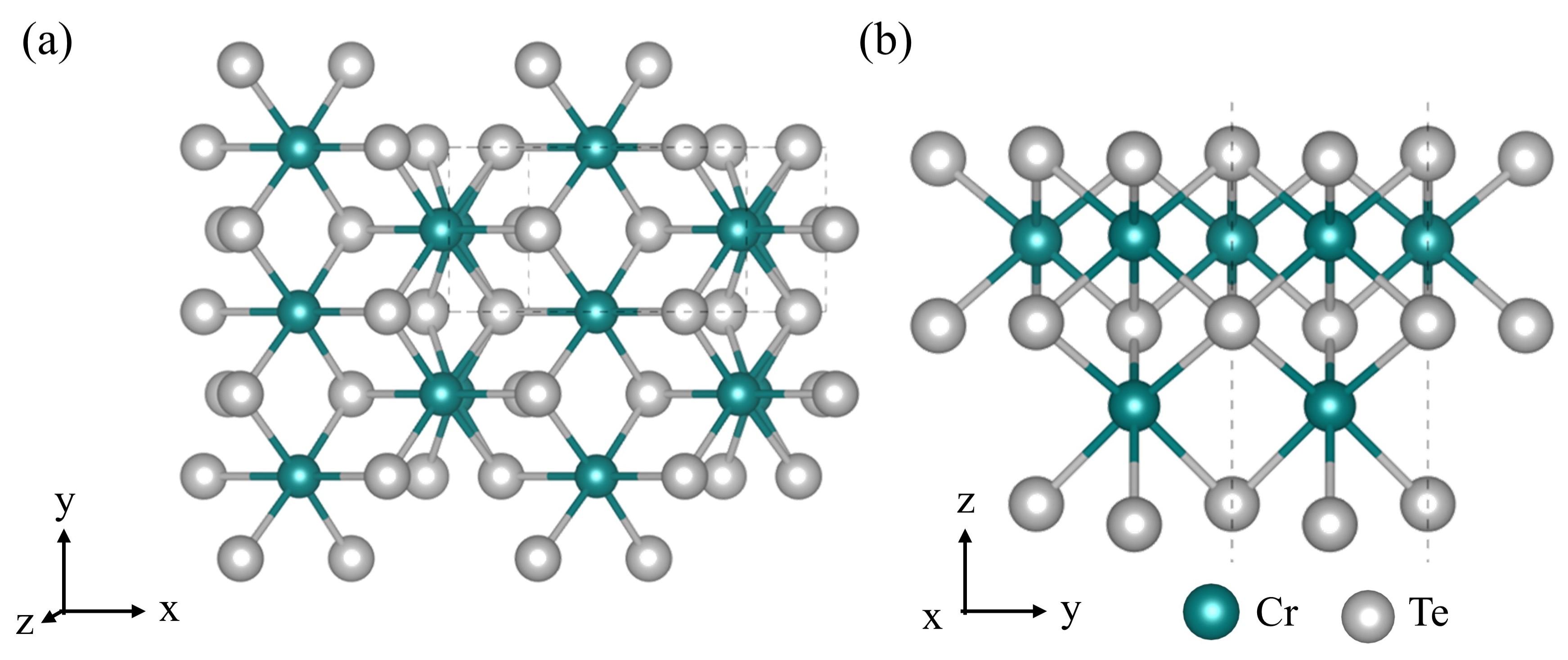}\\
	\caption{Crystal structure of Cr$_3$Te$_6$ . (a) Top view (b) Side view.}\label{fig1}
\end{figure*}

\subsection{Calculate exchange coupling J from superexchange paths}
The crystal structure of 2D Cr$_3$Te$_6$ is shown in Fig. \ref{fig1}, where the space goup is Pm (No.6). In experiment, it is a ferromagnetic metal with high Tc = 344 K \cite{Chua2021}. To theoretically study its magnetic properties, we considered seven different magnetic configurations, including a ferromagnetic (FM) , a ferrimagnetic (FIM), and five antiferromagnetic (AFM) configurations, as discussed in Supplemental Materials \cite{Li}. The calculation results show that the magnetic ground state is ferromagnetic, consistent with the experimental results. Since the superexchange interaction has been suggested to dominate the magnetic interactions in these 2D van der Waals ferromagnetic semiconductors and metals, we study the superexchange interactions in 2D Cr$_3$Te$_6$. 

\begin{figure}[phbpt]
	\centering
	\includegraphics[scale=0.3,angle=0]{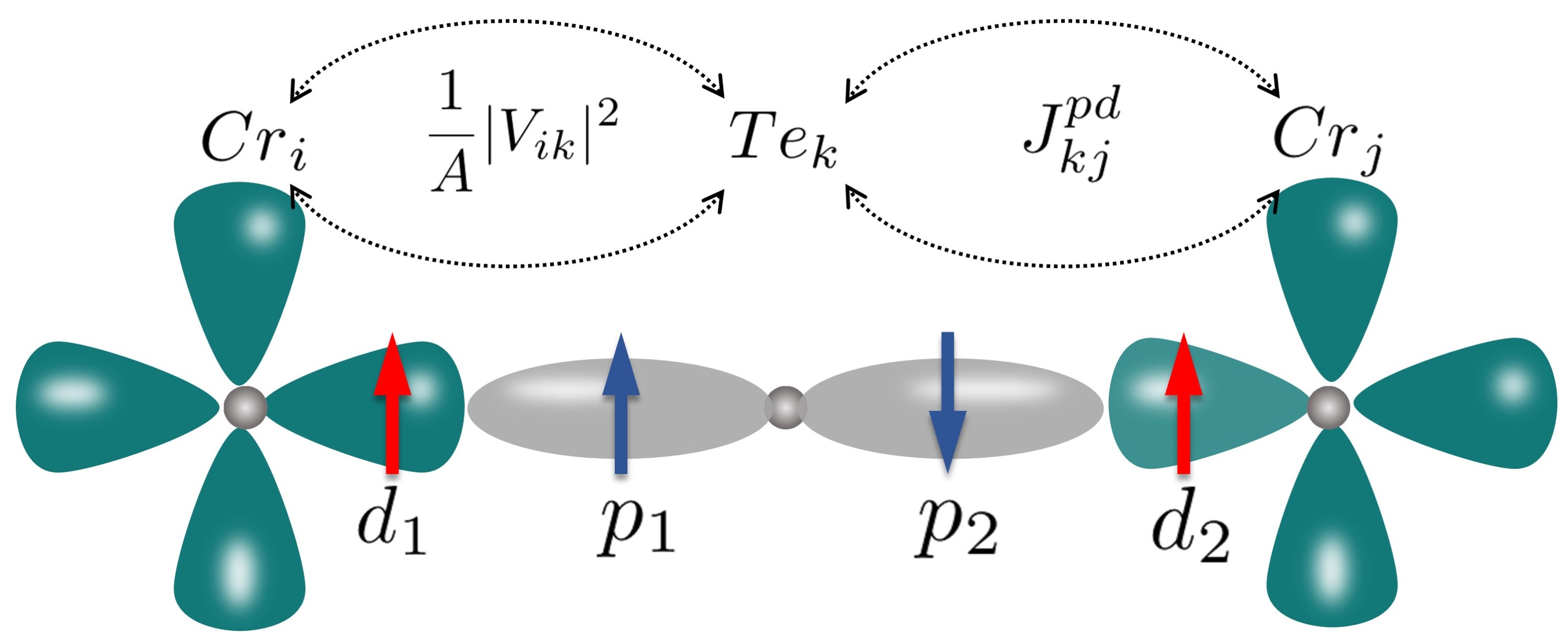}\\
	\caption{Schematic picture of superexchange interaction by a Cr-Te-Cr model. There are two process, one is direct exchange process between Cr$_j$ and Te$_k$, noted as J$_{kj}^{pd}$, and the other is electron hopping between Te$_k$ and Cr$_i$, noted as $|V_{ik}|^2$/A. See text for details.}\label{fig2}
\end{figure}

The superexchange interaction can be reasonably descried by a simple Cr-Te-Cr model \cite{Dai2017}, as shown in Fig. \ref{fig2}. There are two Cr atoms at sites i and j, and one Te atom at site k between the two Cr atoms. By the perturbation calculation, the superexchange coupling J$_{ij}$ between the two Cr atoms can be obtained as \cite{Dai2017},

\begin{equation}
	\begin{aligned}
		J_{ij} =&  (\frac{1}{E_{\uparrow\downarrow}^2}-\frac{1}{E_{\uparrow\uparrow}^2})\sum\limits_{k,p,d}|V_{ik}|^2J_{kj}^{pd}\\
		=& \frac{1}{A}\sum\limits_{k,p,d}|V_{ik}|^2J^{pd}_{kj}.
	\end{aligned}
	\label{eq1}
\end{equation}
The indirect exchange coupling J$_{ij}$ is consisting of two processes. One is the direct exchange process between the d electron of Cr at site j and the p electrons of Te at site k, presented by J$_{kj}^{pd}$. The other is the electron hopping process between p electrons of Te atom at site k and d electrons of Cr atom at site i, presented by |V$_{ik}|^2/A$. V$_{ik}$ is the hopping parameter between d electrons of Cr atom at site i and p electrons of Te atom at site k. Here, A = 1/(1/E$_{\uparrow\downarrow}^2$-1/E$_{\uparrow\uparrow}^2$), and is taken as a pending parameter. E$_{\uparrow\uparrow}$ and E$_{\uparrow\downarrow}$ are energies of two d electrons at Cr atom at site i with parallel and antiparallel spins, respectively. The direct exchange coupling J$_{kj}^{pd}$ can be expressed as \cite{Dong2019,Dong2020,You2020,You2021}: 

\begin{equation}
	\begin{aligned}
		J_{kj}^{pd} = \frac{2|V_{kj}|^2}{|E_{k}^p-E_{j}^d|}.
	\end{aligned}
	\label{eq2}
\end{equation}

V$_{kj}$ is the hopping parameter between p electrons of Te atom at site k and d electrons of Cr atom at site j. E$_k^p$ is the energy of p electrons of Te atom at site k, and E$_j^d$ is the energy of d electrons of Cr atom at site j.

By the DFT and Wannier function calculations, the parameters V$_{ik}$, V$_{kj}$, E$_k^p$, and E$_j^d$ in Eqs. (\ref{eq1}) and (\ref{eq2}) can be calculated. The J$_{ij}$A can be obtained by counting all the possible k sites of Te atoms, p orbitals of Te atoms, and d orbitals of Cr atoms.

\begin{figure*}[bht]
	\centering
	\includegraphics[scale=0.58,angle=0]{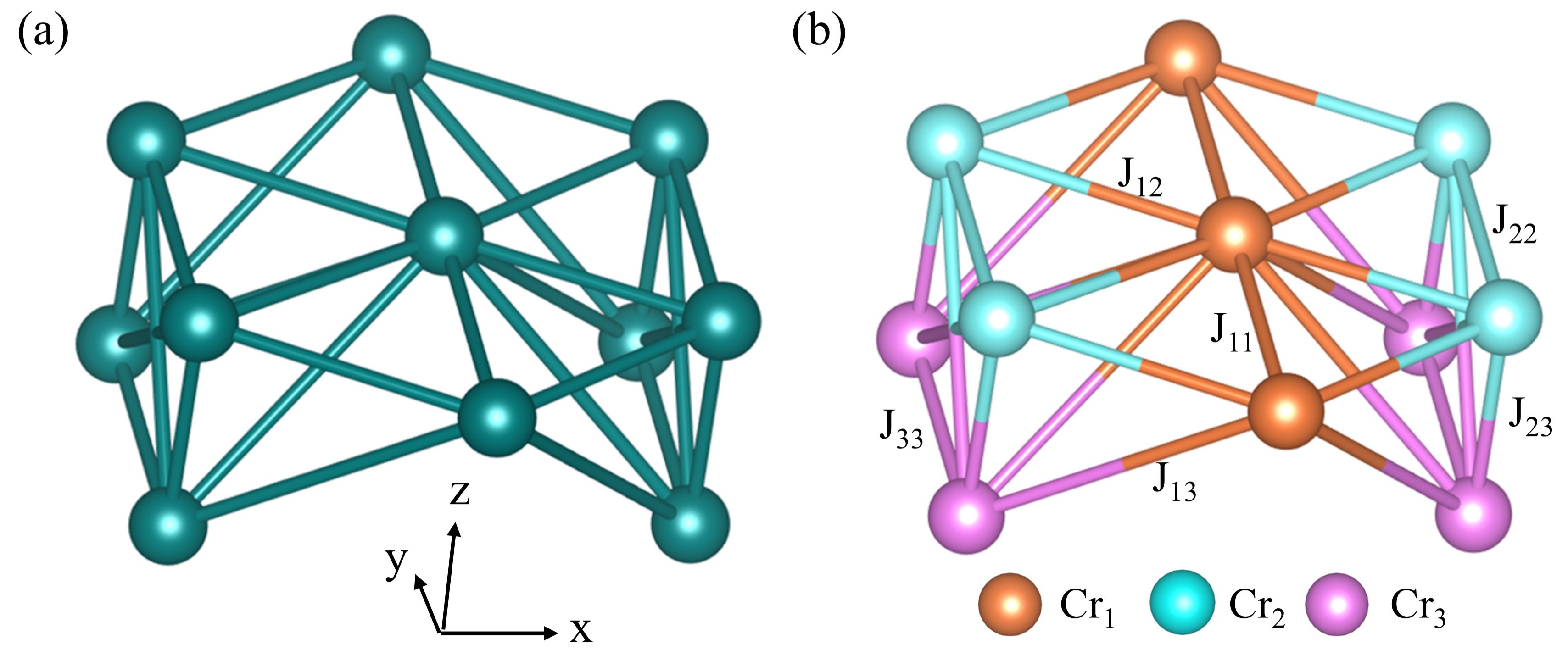}\\
	\caption{(a) The crystal structure of Cr atoms in 2D Cr$_3$Te$_6$. (b) The magnetic structure of Cr atoms in 2D Cr$_3$Te$_6$, calculated by Eqs. (\ref{eq1}) and (\ref{eq2}). }\label{fig3}
\end{figure*}

\begin{table*}[!bhtb]
	\setlength{\tabcolsep}{7mm}{
		\scalebox{1}
		
		\caption{For 2D Cr$_3$Te$_6$, the calculated exchange coupling parameters J$_{ij}$A in Eqs.(\ref{eq1}) and (\ref{eq2}), by the density functional theory and Wannier functional calculations. A is a pending parameter. The unit of J$_{ij}$A is meV$^3$. }
		{
			
			\begin{tabular}{cccccc}
				\hline
				\hline
				J$_{11}$A & 
				J$_{22}$A & 
				J$_{33}$A & 
				J$_{12}$A & 
				J$_{13}$A & 
				J$_{23}$A 
				\\
				\hline
				40&26&53&29&44&83
				\\
				\hline								\hline
			\end{tabular}
		\label{table1}
	}}
	
\end{table*}

\begin{table*}[btbht]
	\setlength{\tabcolsep}{3mm}
		\caption{For 2D magnetic metal Cr$_3$Te$_6$ and semiconductors Cr$_3$O$_6$ and Mn$_3$O$_6$, the parameter A (in unit of meV$^{-2}$) in Eq. (\ref{eq1}), the exchange couping parameters J$_{ij}$S$^2$ and the magnetic anisotropy parameter DS$^2$ (in unit of meV) in the Hamiltonian in Eq. (\ref{eq3}), and the estimated Curie temperature Tc. See text for details.  }
		{
		\scalebox{1}
		
		{
			\begin{tabular}{cccccccccc}
				\hline
				\hline
				Materials&
				A&
				J$_{11}$S$^2$ & 
				J$_{22}$S$^2$ & 
				J$_{33}$S$^2$ & 
				J$_{12}$S$^2$ & 
				J$_{13}$S$^2$ & 
				J$_{23}$S$^2$ & 
				DS$^2$ &
				Tc (K)
				\\
				\hline
				Cr$_3$Te$_6$&-27& -17.1& -11.5&-24.4&-12.6&-19.6&-37.4&-0.14&328\\
				Cr$_3$O$_6$&-36& -18.9& -14.6&-10.1&-18.7&-1.8&-3.1&0.04&218\\
				Mn$_3$O$_6$&-465& -11.9& -7.6&-50.4&-15.9&-5.2&-10.7&-0.09&208\\
				\hline
				\hline
				
			\end{tabular}
	}}

	\label{table2}
\end{table*}


From the calculated results in Table \ref{table1}, it is suggested that there are six different nearest-neighbor couplings, denoted as J$_{11}$, J$_{22}$, J$_{33}$, J$_{12}$, J$_{13}$, and J$_{23}$, as shown in Fig. \ref{fig3}(b). Accordingly, there are three kinds of Cr atoms, noted as Cr$_1$, Cr$_2$, and Cr$_3$. Based on the results in Table \ref{table1}, the effective spin Hamiltonian can be written as 

\begin{equation}
	\begin{aligned}
		H = 
		&J_{11}\sum\limits_{n}\vec{S}_{1n} \cdot \vec{S}_{1n}
		+J_{22}\sum\limits_{n}\vec{S}_{2n} \cdot \vec{S}_{2n}
		+J_{33}\sum\limits_{n}\vec{S}_{3n} \cdot \vec{S}_{3n}\\
		+&J_{12}\sum\limits_{n}\vec{S}_{1n} \cdot \vec{S}_{2n}
		+J_{13}\sum\limits_{n}\vec{S}_{1n} \cdot \vec{S}_{3n}
		+J_{23}\sum\limits_{n}\vec{S}_{2n} \cdot \vec{S}_{3n}\\
		+&D\sum\limits_{n}(S_{1nz}^2+S_{2nz}^2+S_{3nz}^2),
	\end{aligned}
	\label{eq3}
\end{equation}
where J$_{ij}$ means magnetic coupling between Cr$_i$ and Cr$_j$, as indicated in Fig. \ref{fig3}(b). D represents the magnetic anisotropy energy (MAE) of Cr$_3$Te$_6$.

\subsection{Determine the parameters D and A}
The single-ion magnetic anisotropy parameter DS$^2$ can be obtained by: DS$^2$=(E$_{\perp}$-E$_{\parallel}$)/6, where E$_{\perp}$ and E$_{\parallel}$ are energies of Cr$_3$Te$_6$ with out-of-plane and in-plane polarizations in FM state, respectively. It has DS$^2$ = -0.14 meV/Cr for 2D Cr$_3$Te$_6$, which is in agreement with the value of -0.13 meV/Cr reported in the previous study of Cr$_3$Te$_6$ \cite{Chua2021}.

The parameter A can be calculated in the following way. Considering a FM and  an AFM configurations, the total energy of Eq. (\ref{eq3}) without MAE term can be respectively expressed as \cite{Li}:
\begin{eqnarray}    \label{eq4}
	E_{FM}&=&2J_{11}S_1^2+2J_{22}S_2^2+2J_{33}S_3^2+8J_{12}S_1S_2\nonumber    \\
	&\;&+2J_{23}S_2S_3+8J_{13}S_1S_3+E_0 \nonumber\\
		&=&11838/A+E_0, \nonumber\\
	E_{AFM1}&=&2J_{11}S_1^2+2J_{22}S_2^2-2J_{33}S_3^2-8J_{12}S_1S_2+E_0\nonumber\\
	&=&-2502/A+E_0. \nonumber\\
\end{eqnarray}

The results in Table \ref{table1} are used to obtain the final expressions in Eq. (\ref{eq4}).  Since two parameters A and E$_0$ are kept, two spin configurations FM and AFM1 are considered here. Discussion on the choice of spin configurations is given in Supplemental Materials \cite{Li}. For the FM spin configuration, the ground state of Cr$_3$Te$_6$, the total energy is taken as E$_{FM}$ = 0 for the energy reference. The total energy of AFM1, E$_{AFM1}$ = 535 meV is obtained by the DFT calculation. The parameters A and E$_0$ are obtained by solving Eq. (\ref{eq4}), and the six exchange coupling parameters J$_{ij}$ can be obtained by Table \ref{table1}. The results are given in Table  \ref{table2}.

\subsection{Estimate Tc by Monte Carlo simulation }
To calculate the Curie temperature, we used the Monte Carlo program for the Heisenberg-type Hamiltonian in Eq. (\ref{eq3}) with parameters in Table \ref{table2}. The Monte Carlo simulation was performed on a 30$\sqrt{3}$ $\times $30$\sqrt{3}$ lattice with more than 1$\times$10$^6$ steps for each temperature. The first two-third steps were discarded, and the last one-thirds steps were used to calculate the temperature-dependent physical quantities. As shown in Table \ref{table2} and Fig. \ref{fig4} (d), the calculated Tc = 328 K for 2D Cr$_3$Te$_6$, close to the Tc = 344 K of 2D Cr$_3$Te$_6$ in the experiment \cite{Chua2021}. Discussion on the choice of spin configurations and the estimation of exchange couplings J$_{ij}$ and Tc is given in Supplemental Materials \cite{Li}.

\section{P\lowercase{rediction of} T\lowercase{wo} H\lowercase{igh} C\lowercase{urie} T\lowercase{emperature} M\lowercase{agnetic}
S\lowercase{emiconductors} C\lowercase{r}$_3$O$_6$ \lowercase{and} M\lowercase{n}$_3$O$_6$}

\begin{figure}[!bht]
	\centering
	\includegraphics[scale=0.33,angle=0]{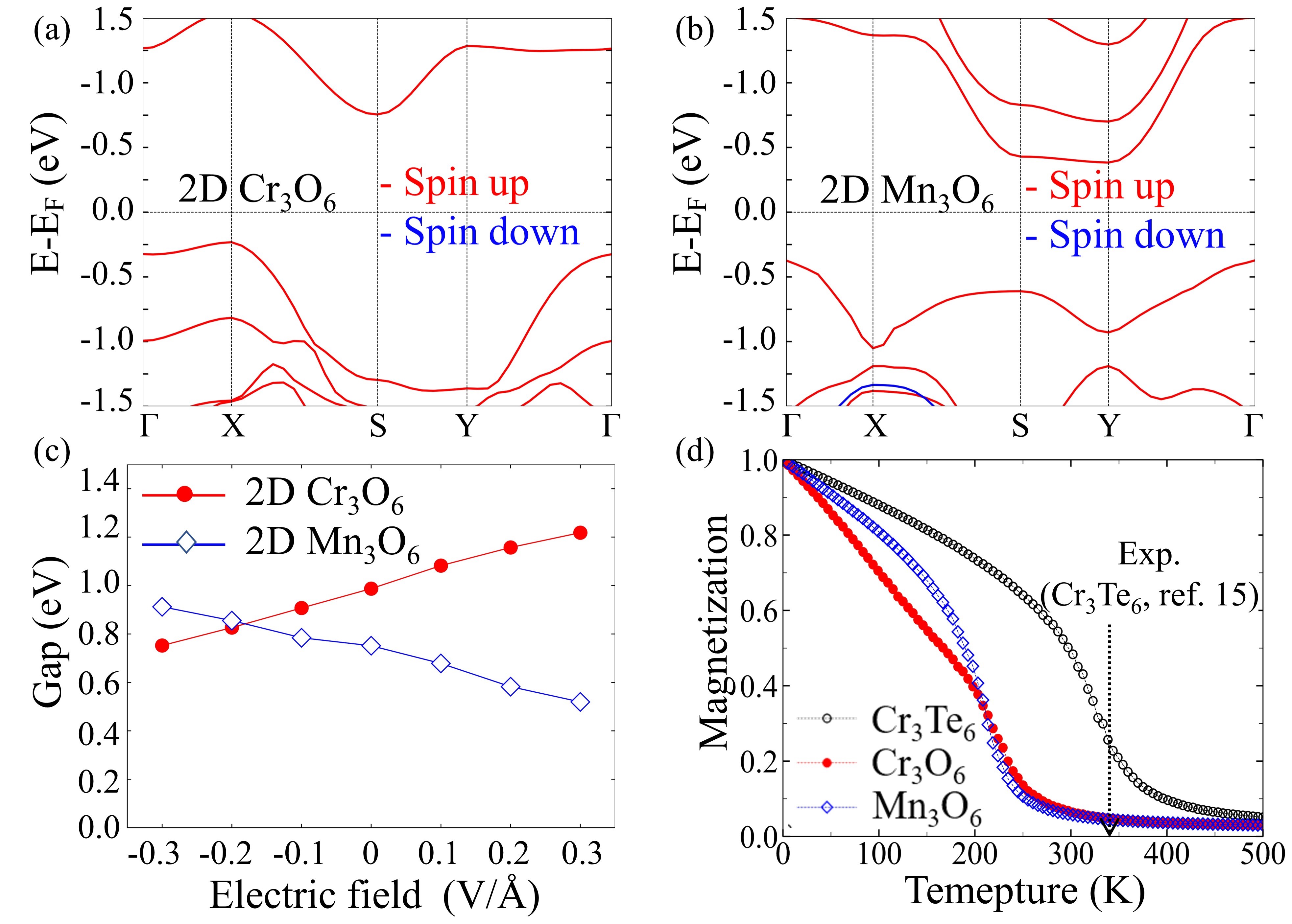}\\
	\caption{(a) Band structures of Cr$_3$O$_6$ with a bandgap of 0.99 eV. (b) Band structures of Mn$_3$O$_6$ with a bandgap of 0.75 eV. (c) Energy gap of Cr$_3$O$_6$ and Mn$_3$O$_6$ under external electric field out-plane. (d) The magnetic moment of Cr$_3$Te$_6$, Cr$_3$O$_6$, and Mn$_3$O$_6$ varies with temperature. }\label{fig4}
\end{figure}

Inspired by the high Tc in the 2D magnetic metal Cr$_3$Te$_6$, we explore the possible high Tc magnetic semiconductors with the same crystal structure of Cr$_3$Te$_6$ by the DFT calculations. We obtain two stable ferromagnetic semiconductors Cr$_3$O$_6$ and Mn$_3$O$_6$. In order to study the stability of the 2D Cr$_3$O$_6$ and Mn$_3$O$_6$, we calculate the phonon spectrum. As shown in Supplemental Materials \cite{Li}, there is no imaginary frequency, indicating the dynamical stability. In addition, we performed molecular dynamics simulations of Cr$_3$O$_6$ and Mn$_3$O$_6$ at 250 K, taking the NVT ensemble (constant temperature and volume) and run for 6 ps. The results show that 2D Cr$_3$O$_6$ and Mn$_3$O$_6$ are thermodynamically stable \cite{Li}. These calculation results suggest that 2D Cr$_3$O$_6$ and Mn$_3$O$_6$ may be feasible in experiment.

The band structure of 2D Cr$_3$O$_6$ and Mn$_3$O$_6$ is shown in Figs. 4(a) and 4(b), respectively, where the band gap is 0.99 eV for Cr$_3$O$_6$ and 0.75 eV for Mn$_3$O$_6$. As shown in Figs. \ref{fig4}(a) and (b), the band gap for 2D Cr$_3$O$_6$ and Mn$_3$O$_6$ is 0.99 eV and 0.75 eV, respectively. When applying an out-of-plane electric field with a range of $\pm$ 0.3 V/Å, which is possible in experiment \cite{Domaretskiy_2022}, the band gap of Cr$_3$O$_6$ (Mn$_3$O$_6$) increases (decreases) with increasing electric field, as shown in Fig. \ref{fig4}(c). By the same calculation method above, the parameter A, the similar Heisenberg models in Eq. \ref{eq3} with six nearest-neighbor exchange coupling J$_{ij}$ are obtained for the 2D Cr$_3$O$_6$ and Mn$_3$O$_6$. The parameters A, J$_{ij}$ and D are calculated and shown in Table \ref{table2}. The spin polarization of Cr$_3$O$_6$ and Mn$_3$O$_6$ is in-plane (DS$^2$ = 0.04 meV) and out-of-plane (DS$^2$ = -0.09 meV), respectively. Fig. \ref{fig4}(d) shows the magnetization as a function of temperature for 2D Cr$_3$Te$_6$, Cr$_3$O$_6$ and Mn$_3$O$_6$. The calculated Curie temperature is Tc = 218 K for 2D Cr$_3$O$_6$ and Tc = 208 K for 2D Mn$_3$O$_6$, respectively.

\section{Conclusion}
	Based on the DFT and Wannier function calculations, we propose a method for constructing the 2D Heisenberg model with the superexchange interactions. By this method, we obtain a 2D Heisenberg model with six different nearest-neighbor exchange couplings for the 2D ferromagnetic metal Cr$_3$Te$_6$. The calculated Curie temperature Tc = 328 K is close to the Tc = 344 K of Cr$_3$Te$_6$ in the experiment. In addition, we predicted two 2D magnetic semiconductors: Cr$_3$O$_6$ with band gap of 0.99 eV and Tc = 218 K, and Mn$_3$O$_6$ with band gap of 0.75 eV and Tc = 208 K, where the similar 2D Heisenberg models are obtained. The complex Heisenberg model developed from the simple crystal structure shows the power of our method to study the magnetic properties in these 2D magnetic metals and semiconductors.

\section* {Acknowledgements}
	This work is supported in part by the National Natural Science Foundation of China (Grants No. 12074378 and No. 11834014), the Beijing Natural Science Foundation (Grant No. Z190011), the National Key R\&D Program of China  (Grant No. 2018YFA0305800), the Beijing Municipal Science and Technology Commission (Grant No. Z191100007219013), the Chinese Academy of Sciences (Grants No. YSBR-030 and No. Y929013EA2), and the Strategic Priority Research Program of Chinese Academy of Sciences (Grants No. XDB28000000 and No. XDB33000000).

\bibliographystyle{apsrev4-2}
\bibliography{ref0}

\end{document}